\documentclass[letterpaper]{article} 
\usepackage{aaai25}  
\usepackage{times}  
\usepackage{helvet}  
\usepackage{courier}  
\usepackage[hyphens]{url}  
\usepackage{graphicx} 
\usepackage{natbib}  
\usepackage{caption} 
\usepackage{algorithm}
\usepackage{algorithmic}
\usepackage{multirow}
\usepackage{booktabs}
\usepackage{amsmath}
\usepackage{amsfonts}
\usepackage{newfloat}
\usepackage{listings}
\DeclareCaptionStyle{ruled}{labelfont=normalfont,labelsep=colon,strut=off} 
\floatstyle{ruled}
\newfloat{listing}{tb}{lst}{}
\floatname{listing}{Listing}
\title{Predicting User Behavior in Smart Spaces with LLM-Enhanced Logs \\
and Personalized Prompts}
\author{
    Yunpeng Song,
    Jiawei Li,
    Yiheng Bian,
    Zhongmin Cai\footnote{Corresponding author.}
}
\affiliations{
    MOE KLINNS Lab, Xi’an Jiaotong University, Xi'an, China\\    

    yunpengs@xjtu.edu.cn, zmcai@sei.xjtu.edu.cn
}

\usepackage{bibentry}

\begin{document}

\maketitle

\begin{abstract}
Enhancing the intelligence of smart systems, such as smart homes, smart vehicles, and smart grids, critically depends on developing sophisticated planning capabilities that can anticipate the next desired function based on historical interactions. While existing methods view user behaviors as sequential data and apply models like RNNs and Transformers to predict future actions, they often fail to incorporate domain knowledge and capture personalized user preferences. In this paper, we propose a novel approach that incorporates LLM-enhanced logs and personalized prompts. Our approach first constructs a graph that captures individual behavior preferences derived from their interaction histories. This graph effectively transforms into a soft continuous prompt that precedes the sequence of user behaviors. Then our approach leverages the vast general knowledge and robust reasoning capabilities of a pretrained LLM to enrich the oversimplified and incomplete log records. By enhancing these logs semantically, our approach better understands the user's actions and intentions, especially for those rare events in the dataset. We evaluate the method across four real-world datasets from both smart vehicle and smart home settings. The findings validate the effectiveness of our LLM-enhanced description and personalized prompt, shedding light on potential ways to advance the intelligence of smart space.
\end{abstract}

%

\section{Introduction}


In today's society, smart devices have become an indispensable part of our daily lives, significantly enriching our life experiences. However, despite the variety of functionalities these smart devices offer, they largely operate independently of one another, lacking deep inter-connectivity. This isolation hinders the formation of truly ``smart'' systems. Take the smart vehicle scenario as an example: on a cold workday morning, a driver might open their car door, followed by the need to adjust the seat position, turn on the heating, set the navigation to plan a congestion-free route, and play their favorite music. While these functions are straightforward, manually executing each one can be tedious. Enabling a simple action, like opening a door, to automatically trigger all related tasks would significantly enhance the user experience in smart settings. Simplifying the interaction between users and smart devices to allow the devices to understand and automatically adapt to users' needs in various scenarios represents an urgent yet challenging goal~\cite{desolda2017empowering}.

This vision intersects with two main areas of research: human activity prediction and sequential recommendation. In human activity prediction, previous works in the context of smart space attempt to build models from historical interaction data between users and their environments, with the goal of automating the recognition of users' intentions and planning the next move~\cite{tax2018human,maharjan2019development,mohamed2022future,dunne2023semantic}. For instance, deep learning models have been employed to analyze various sensor data within smart homes to forecast subsequent user activities~\cite{krishna2018lstm}. On the other hand, sequential recommendation focuses on the online activity context suggesting content that users may find engaging based on their behaviors~\cite{hou2022towards,li2023text}. This includes analyzing sequences of user online interactions, such as movie watching, web browsing, clicking, bookmarking, and purchasing, to tailor recommendations of potentially interesting content, like movies or products.

However, in real-world smart spaces, existing methods face challenges that differ markedly from those in recommendation systems. While recommendation systems can access extensive side information, which includes the names, brands, and attributes of purchased items~\cite{li2023text}, and the content of movies that interest users, data logs from devices and sensors in smart spaces are often sparse or even incomplete. Consequently, earlier methods that employed one-hot encoding to represent different events as unrelated IDs frequently missed the inherent connections between related actions. For example, opening a refrigerator and preheating an oven are both linked to the intention of eating, yet are treated as unrelated by one-hot encoding. This approach fails to recognize the natural associations between activities, leading to a disconnect in representations, particularly for low-frequency actions that are poorly represented. 

Additionally, due to the limited number of users in previous smart space datasets~\cite{tax2018human,park2022enhanced,wang2022predicting,dunne2023semantic}, prior research in smart spaces typically employs a single model to predict a range of user actions, neglecting the individual preferences that result in behavioral differences. For instance, the routine of going to sleep could involve actions like drawing the curtains and switching off the lights for some, while for others, it might include playing soft music and setting an alarm. This variation highlights the inadequacy of a one-size-fits-all approach in predicting user actions without considering individual preferences, which is crucial for effective smart space automation.

To address the challenges in predicting user behavior within smart spaces, we propose a novel approach that incorporates LLM-enhanced logs and personalized prompts. This method taps into the vast general knowledge and robust reasoning capabilities of a pretrained LLM, using in-context learning to enrich the oversimplified and incomplete log records. By enhancing these logs semantically, we can better understand the user's actions and intentions. Additionally, our approach constructs a graph that captures individual behavior preferences derived from their interaction histories. This graph effectively transforms into a soft continuous prompt that precedes the sequence of user behaviors. Through prompt tuning, the model is guided to make varied predictions about user behavior based on this prompt.

We primarily evaluate our approach through experiments conducted on two large-scale, real-world datasets of user interactions with smart vehicles, complemented by two additional datasets focusing on smart home settings. The experimental outcomes demonstrate the superior performance of our approach when compared to state-of-the-art methods in terms of predicting a user's next move within smart vehicle contexts. By incorporating LLMs for description augmentation, our approach significantly boosts performance in scenarios involving low-frequency events. Furthermore, our framework demonstrates robust adaptability, effectively handling cold-start scenarios with new users and seamlessly transitioning to smart home contexts. The contributions are summarized as follows: 

\begin{itemize}
    \item We introduce a novel prompt-based approach for predicting user behavior, which adapts dynamically to individual user preferences, making it a pioneering application of prompt tuning for customized predictions within the smart space context.    
    \item Utilizing a pre-trained LLM, our method significantly enhances event representations by augmenting them with semantic descriptions, particularly effective for addressing rare events.    
    \item We validate our approach with extensive experiments on datasets from two different domains, demonstrating its effectiveness in accurately predicting user behavior. 
\end{itemize}

\begin{figure*}
    \centering
    \includegraphics[width=1\linewidth]{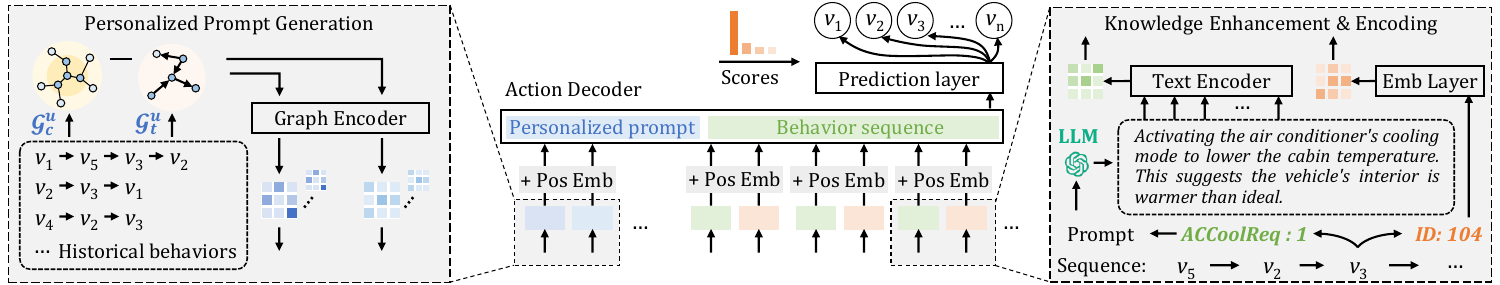}    
    \caption{Overview of the proposed method for smart space interaction prediction. Our method constructs graphs from the user's historical sequences to model individual preferences and generate personalized prompts. It also augments event descriptions using an LLM's common knowledge.}
    \label{fig:overview}
\end{figure*}

\section{Related Work}
This section provides an overview of the existing research in three main areas: human activity prediction within smart environments, sequential recommendation systems for applications such as movies, e-commerce, and other services, and prompt tuning for recommendations.

\subsection{Human Activity Prediction}

Human activity prediction in smart environments aims to enhance the user experience by reducing the need for direct interaction and enabling more proactive support~\cite{minor2017forecasting,tax2018human,almeida2018predicting,yang2020multi}. Research in this domain has primarily leveraged data-driven approaches, employing machine learning algorithms to analyze sensor-generated event logs that capture daily user interactions. These logs are annotated with activity labels to train prediction models. 
Early efforts~\cite{minor2015data, minor2017learning} utilized simple predictors with manually crafted features for prediction given sequences of past events. Advancements in the field have introduced sophisticated models like LSTM and GRU networks for discrete sequence modeling~\cite{tax2018human,krishna2018lstm}. Alternative methods employing CNN~\cite{wang2022predicting} and DNN~\cite{park2022enhanced} have been proposed to extract local features from the status of all sensors, using LSTM networks for the next activity prediction. Song et al.~\cite{song2024learning} further employed a Transformer decoder to incorporate the multi-source data for generating automation routines. However, these studies primarily overlook the semantic associations between signals and the variations in user behavior preferences by utilizing ID-based methods for modeling events and user profiles. In contrast, our approach enhances signal semantic descriptions using a pre-trained LLM and employs a prompt-based method to distinctively represent individual user behavior preferences.

\subsection{Sequential Recommendation}
Sequential recommendation systems are designed to anticipate a user's next point of interest, which could range from movies to products, by analyzing their past interactions~\cite{wang2022target,hou2023learning,rajput2023recommender,dong2024prompt}. Early research in this field utilized Markov Chains to understand the sequence of item transitions~\cite{shani2005mdp}. However, the integration of deep learning has led to the development of more advanced techniques. One of the first significant contributions in this area was GRU4Rec~\cite{hidasi2015session}, which introduced RNNs to the sequential recommendation domain. SASRec~\cite{kang2018self} and Bert4Rec~\cite{sun2019bert4rec} applied transformers equipped with self-attention mechanisms to capture the dynamic interplay between items. Fastformer~\cite{wu2021fastformer} utilizes additive attention to achieve linear complexity, accelerating the Transformer model. DuoRec~\cite{qiu2022contrastive} and CL4SRec~\cite{xie2022contrastive} explored contrastive learning to refine the process of learning item embeddings. More recently, UniSRec~\cite{hou2022towards} and RecFormer~\cite{li2023text} incorporate textual attributes of products into models to enrich the representations of items further. However, these studies primarily concentrate on online behaviors, utilizing datasets with relatively uniform behavioral objectives
such as Amazon Product\footnote{\url{https://cseweb.ucsd.edu/~jmcauley/datasets/amazon_v2}} for purchasing or browsing items or MovieLens\footnote{\url{https://grouplens.org/datasets/movielens}} for watching movies, which differ significantly from our focus. Our research targets predicting actions within the context of smart spaces, encompassing a broader range of activities and involving more extended intentions.

\subsection{Prompt Tuning}
Prompt tuning is originally developed in the field of natural language processing, aiming to tailor input to help pretrained language models adapt to different downstream tasks. This approach has recently been employed for recommendation~\cite{wang2023plate,li2023personalized,wu2024personalized,yang2024empirical,dong2024prompt}. For instance, P5~\cite{geng2022recommendation} integrated user–item information and corresponding features with personalized hard prompts to form a unified and instruction-based method for varied recommendation tasks. PEPLER~\cite{li2023personalized} utilized item features and ID embeddings to generate recommendation explanations. PPR~\cite{wu2024personalized} built personalized soft prompt via a prompt generator based on user profiles for cold-start recommendation. However, while most existing methods depend on side information, such as user profiles like age and gender, to represent user preferences, our approach leverages historical interaction data to construct graphs capturing personalized user behaviors without relying on user profiles. 

\section{Method}

\subsection{Approach Overview}
\subsubsection{Problem Setup.}

In this study, we aim to predict a user's next action by analyzing their past interactions, framing this task as a multi-class classification challenge. We introduce the notion of a user set, denoted by $\mathcal{U}$, and an event set, $\mathcal{V}$. For a given user $u$, one of their interaction sequence is represented as $S_u = \{v_1, v_2, ..., v_{|S_u|}\}$, where $|S_u|$ donates the sequence length. Each event $v_t \in \mathcal{V}$ in this sequence, consisting of both the event name and its value, is identified by a unique ID. The event name, typically a brief, technical descriptor like ``FLWdwClsReq'' for ``front left door close request'', records the function of the event. Each event can have different values indicating various user actions and lead to different IDs; typically, a value of 0 signifies deactivation, while 1 indicates activation. Our objective is to leverage such historical sequences to forecast the user's next event, aiming to propose a list of the most probable future events ranked by their likelihood. This prediction is formulated as:
\[
v_{|S_u|+1} = \text{argmax}_{v \in \mathcal{V}} P(v | S_u),
\]
where $P(v | S_u)$ denotes the probability of event $v$ being the next action given the sequence of past events $S_u$.

\subsubsection{Framework Overview.}
Smart spaces encompass a variety of functionalities, leading to diverse user behaviors within these settings. Understanding user intentions and predicting their actions to achieve a higher degree of automation is essential yet challenging. One major difficulty is that the log data of user interactions are often too simplistic or incomplete. For instance, entries like ``ExCrtLtAutoCfg: 0'' provide little insight into the semantics of the actions they represent. And there is significant variability in user behavior preferences; even with a single intent, different users follow varied operational sequences based on personal habits. Another challenge is the long-tail distribution of feature usage. Many functions are rarely used, which complicates the learning process for predictive models, as these underutilized features do not provide enough data to train effectively.

To address these challenges, we propose a data-driven approach that incorporates LLM-
enhanced logs and personalized prompts. Initially, our approach constructs two graphs based on the user’s historical interactions to capture individual preferences and generate personalized prompts. These prompts are inserted at the beginning of user behavior sequences to guide the model in making differentiated predictions tailored to individual characteristics. Then our approach uses an LLM to enrich event descriptions with detailed natural language explanations. For example, the code ``ExCrtLtAutoCfg: 0'' is expanded to clarify that it represents ``disabling the automatic activation feature for the exterior courtesy lights.'' This enhancement preserves the semantic coherence among events in the embedding space, which is crucial for forming robust representations, especially for less frequent events. Additionally, a decoder then processes these inputs to predict the user's next likely action. The overall architecture of our framework is depicted in Figure~\ref{fig:overview}.

\subsection{Personalized Prompt Generator}
Prompt tuning is typically used to bridge the gap between a pre-trained model's objectives and the specific requirements of a downstream task~\cite{brown2020language}. Here, we employ prompt tuning to craft personalized prompts that integrate individual users' behavioral patterns directly into the model. This customization enables the model to make predictions that truly reflect user-specific characteristics. NLP research has relied on hard prompt designing~\cite{geng2022recommendation}, which poses two main limitations for our application~\cite{yang2024empirical}: first, unlike words in NLP, the tokens (event IDs) in our task lack clear, inherent meanings. Second, recommendation requires personalization, meaning prompts must be tailored to individual users. Considering the vast number of users in real-world systems, manually creating custom prompts for each user is infeasible.

To address the challenges, we have adopted a soft prompt based method by building interaction graphs and converting them into continuous prompts to reflect the user-level preferences. To effectively capture the distinct behavioral patterns of different users, we first construct a base graph $\mathcal{G} = (\mathcal{V}, \mathcal{E})$ using historical event sequences. In this weighted directed graph, $\mathcal{V}$ represents all the events, and $\mathcal{E} = {e_{ij}}$, where each edge $e_{ij}$ is weighted by the frequency of event $v_i$ being followed by event $v_j$ in the historical data. Building on this foundation, we then create personalized transition and co-occurrence graphs for each user to illustrate the dependencies among events in a user's behavior sequence.

The transition graph, denoted as $\mathcal{G}_t$, is a weighted directed graph designed to capture the specific sequential relationships between events. For user $u$, adjacent events $v_i$ and $v_j$ in the sequence are connected by an edge $e_{ij}^t$ in $\mathcal{G}_t^u$ if the following conditions are met:
\[w_{ij} > (1+\lambda) w_{ji}\]

Here, $w_{ij}$ is the weight of the edge from $v_i$ to $v_j$, and $w_{ji}$ is the weight of the reverse edge. The parameter $\lambda \geq 0$ differentiates between transitional relationships (where one event significantly leads to another) and co-occurrence relationships (where events happen concurrently). In contrast, the co-occurrence graph, $\mathcal{G}_c$, serves a different function. It is a weighted undirected graph that identifies events likely to co-exist within the same sequences but without any implied order. An edge $e_{ij}^c$ exists between $v_i$ and $v_j$ in $\mathcal{G}_c^u$ if 
\[\max(w_{ij}, w_{ji}) \leq (1+\lambda) \min(w_{ij}, w_{ji})\]

After constructing the transition graph $\mathcal{G}_t$ and the co-occurrence graph $\mathcal{G}_o$, we apply a graph convolutional neural network (GCN) to project the representations of events into the prompt embedding space. The GCN utilizes the following message passing method:

\[
\scriptstyle X^{(l+1)}_t = D_t^{-\frac{1}{2}} A_{\mathcal{G}_t^u} D_t^{-\frac{1}{2}} X^{(l)}_t W^{(l)}_t; \quad X^{(l+1)}_c = D_c^{-\frac{1}{2}} A_{\mathcal{G}_c^u} D_c^{-\frac{1}{2}} X^{(l)}_c W^{(l)}_c
\]

Here, $X^{(l)}_t$ and $X^{(l)}_c$ represent the embedding matrices of events over the transition and co-occurrence graphs, respectively, at the \(l\)-th graph layer. $D_t$ and $D_c$ are degree matrices that normalize the graphs to ensure stable training and effective feature propagation. To further transform the embeddings into prompts, we utilize a two-layer perceptron $f(\cdot)$ to process $X^u_t$ and $X^u_c$:
\[
P^u_t = f(X^u_t) = W_2 \cdot \sigma(W_1 \cdot X^u_t + b_1) + b_2
\]
where $W_1$ and $W_2$ are the learnable weights, $b_1$ and $b_2$ are biases, and $\sigma$ denotes the activation function. 

The learned prompts act as prefixes, linked to the user behavior sequence as $S_u=\{p^u_{t_1},p^u_{c_1},\dots,v^u_1,v^u_2,\dots,v^u_{|S_u|}\}$. This arrangement enhances the input representations by incorporating user-specific context. It is important to note that based on the outputs from the GCN, both \(P^u_t\) and \(P^u_c\) may include multiple entries.

\subsection{Enhancing Logs with LLM}

\begin{figure}
    \centering
    \includegraphics[width=1\linewidth]{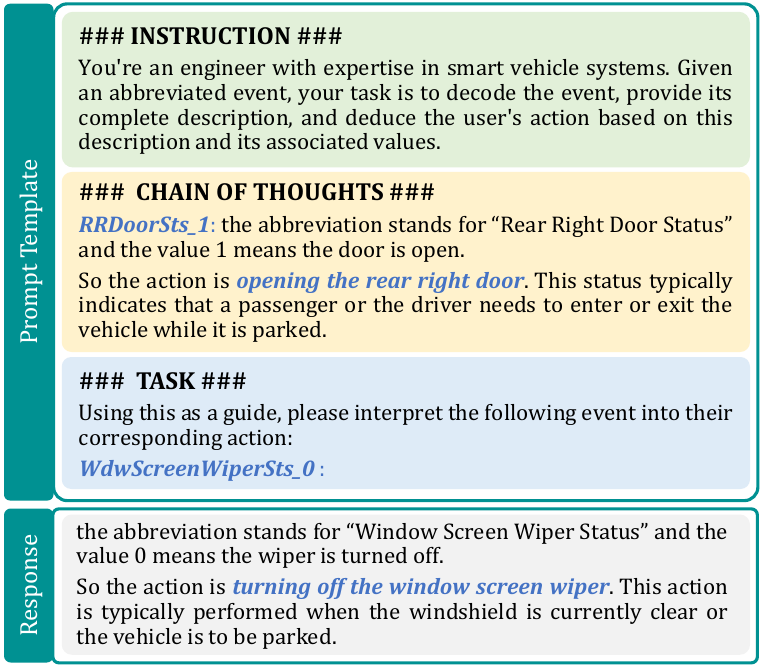}
    \caption{Enhancing an event log with LLM.}
    \label{fig:llm_aug}
\end{figure}

Sensor event logs in smart space are commonly recorded for monitoring or debugging purposes~\cite{cook2009ambient}. These logs essentially reflect user behavior and can effectively assist in understanding user intentions. However, these logs are typically concise. For instance, using ``FLWwClsReq'' to represent ``front left window close request'' in our user-vehicle interaction dataset. This is because they are often reviewed and processed by professionals from service providers. Traditional methods tend to overlook the incomplete textual descriptions of events, opting instead to assign different IDs to recorded log signals to create event embedding tables for encoding~\cite{tax2018human,park2022enhanced,wang2022predicting}. However, by reconstructing these incomplete textual descriptions, we can more accurately depict the relationships between events from a semantic standpoint, rather than merely relying on their sequence of occurrence. This is particularly useful for long-tail infrequent events. For example, even if the event ``RLWwClsReq'' appears rarely and is hard to learn for a model, we can deduce its close semantic similarity to ``FLWwClsReq'' based on their textual information.

We leverage the common sense and reasoning capabilities of a pre-trained LLM to enhance the abbreviated textual descriptions of log events. The training corpora for these models often include a vast array of codes, which feature similar signal abbreviations and annotated descriptions explaining their meanings. This rich dataset enables LLMs to accurately interpret the semantics of various code abbreviations found in logs. Specifically, we set the LLM's role as an expert engineer in intelligent automotive log analysis. Using a Chain of Thought approach~\cite{brown2020language}, we construct prompts that enable the LLM to output possible user actions based on the event logs. Given an event with a description and value, such as ``RRDoorSts: 1'', the LLM reconstructs a more complete description of the event, like ``rear right door status''. Subsequently, it infers user actions based on the description and value; for example, ``1'' indicates the door is open, leading to the inference that the user action is opening the rear right door. The process of prompting is illustrated in Fig.~\ref{fig:llm_aug}. We employed both proprietary Ernie Bot and the open-source Qwen2.5-32B as the LLM to enhance the textual descriptions of all events (description-value pairs) in our datasets, thereby gaining a deeper understanding of the behavioral semantics behind these events.

After the LLM augmentation, each event is associated with a multi-word natural language description. Therefore, we introduce an encoder model (BERT \cite{devlin2018bert}) to learn and utilize these enhanced natural language descriptions for forming representational vectors of events. Given an event with an LLM-enhanced description \(\{d_i^1, d_i^2, \cdots, d_i^m\}\) (where \(m\) is the number of words in the description), we treat this as a sentence and add a special token [CLS] at the beginning. Consequently, the input to the encoder model is formulated as:
\[
    T = \{ [\text{CLS}], D_i^1, D_i^2, \cdots, D_i^m \}
\]
Utilizing BERT's bidirectional transformer architecture, the [CLS] token integrates information from all words in the description and can be used as the textual representation ($t_i$) of the augmented event.

\subsection{Sequence Decoding}

We treat event prediction as a decoding task with personalized prompts, where given the representations of events in a sequence, a sequence decoder interprets user intent and captures behavioral patterns to predict the next move. In this process, we employ a transformer decoder endowed with self-attention mechanisms to aggregate sequence information~\cite{radford2019language}. The decoder comprises several blocks; each block includes a masked multi-headed self-attention layer (MHAttn(·)) and point-wise feed-forward networks (FFN(·)). The use of masks ensures that each prediction is influenced only by prior known outputs. The decoder processes inputs by summing the learned embeddings ($v_i$) and position embeddings ($p_j$) at position $j$. The computation evolves as follows:
\[
    f_j^0 = v_i + p_j
\]
\[
    F^{l+1} = \text{FFN}(\text{MHAttn}(F^l))
\]
Here, $F^l=[f^l_0;\dots;f^l_n]$ represents the $l$-th layer's position-specific representations. We derive the sequence representation from the final hidden state $f_n^L$ at the $n$-th position, across a total of $L$ layers in the decoder.

To effectively differentiate between likely and unlikely subsequent events in a fine-grained manner, we employ a pairwise ranking loss (BPR loss) to optimize the decoder~\cite{rendle2012bpr}. Considering a user's historical event sequence \( S_u = \{v_{1}, v_{2}, \dots, v_{m-1}\} \), the positive sample \( i \) is the target next event, namely \( v_{m} \). The negative sample \( j \) is randomly selected from events not appearing next in the user's interaction data following \( v_{m-1} \). Thus, the ranking loss function is defined as follows:
\[
    \mathcal{L} = -\sum_{j=1}^B \log \sigma(y_{m}^i - y_{m}^j)
\]
Here, \(\sigma\) denotes the sigmoid activation function, and \( y_{m}^i \) and \( y_{m}^j \) represent the likelihood scores of the user \( u \)'s next event being \( i \) and \( j \), respectively, given \( S_u \). The events with the highest scores will be output as the top-k candidates in the model's prediction list. 

\section{Evaluation}

In this section, we outline the experimental setup, followed by the presentation of results and analysis in response to the following research questions.\\
\textbf{RQ1}: Does our method outperform the SOTA methods? \\
\textbf{RQ2}: What is the impact of varying parts in the framework?\\
\textbf{RQ3}: Does our method improve rare event performance?\\
\textbf{RQ4}: How do hyper-parameters affect model performance?


\setlength{\tabcolsep}{1mm}
\begin{table*}
    \small
    \renewcommand{\arraystretch}{1.0}
    \centering    
    \begin{tabular}{c|c|c|cccc|ccccc}
        \toprule
        Sources & Datasets & Metrics & CasLSTM & DnnLSTM & CovLMLC & ASGen & BERT4Rec & DuoRec & CL4SRec & RecFormer & Ours\\       
        \midrule
        \multirow{6}{*}{Industrial}
        & \multirow{3}{*}{City A}
        & H@1   & 0.1987    & 0.1794    & 0.2231    & 0.3018    & 0.2715    & 0.3308    & 0.3286    & 0.3191    & \textbf{0.3450*} \\ 
        && H@3  & 0.3413    & 0.3075    & 0.3670    & 0.4852    & 0.4356    & 0.5114    & 0.5228    & 0.5159    & \textbf{0.5529*} \\
        && N@3  & 0.2817    & 0.2540    & 0.3068    & 0.4065    & 0.3629    & 0.4358    & 0.4427    & 0.4324    & \textbf{0.4596*} \\ 
        \cmidrule{2-12}
        & \multirow{3}{*}{City B}
        & H@1   & 0.2188	& 0.2043	& 0.2286	& 0.3453	& 0.3044	& 0.3497	& 0.3517	& 0.3230	& \textbf{0.3743*} \\ 
        && H@3  & 0.3808	& 0.3543	& 0.4041	& 0.5270	& 0.4567	& 0.5315	& 0.5661	& 0.5387	& \textbf{0.5904*} \\ 
        && N@3  & 0.2974	& 0.2864	& 0.3225	& 0.4640	& 0.4256	& 0.4511	& 0.4589	& 0.4389	& \textbf{0.4925*} \\           

        \midrule
        \multirow{6}{*}{Public}
        & \multirow{3}{*}{Home A}
        & H@1   & 0.2684	& 0.2478	& 0.2737	& 0.3598	& 0.3508	& 0.4015	& 0.3691	& 0.4103	& \textbf{0.4293*} \\ 
        && H@3  & 0.4263	& 0.4094	& 0.4383	& 0.5390	& 0.5298	& 0.5634	& 0.5287	& 0.5731	& \textbf{0.5974*} \\
        && N@3  & 0.3349	& 0.3339	& 0.3573	& 0.4685	& 0.4515	& 0.4973	& 0.4688	& 0.5093	& \textbf{0.5286*} \\       
        \cmidrule{2-12}
        & \multirow{3}{*}{Home B}
        & H@1   & 0.1842	& 0.1954	& 0.2032	& 0.3071	& 0.2878	& 0.3026	& 0.3145	& 0.3118	& \textbf{0.3352*} \\ 
        && H@3  & 0.3497    & 0.3517    & 0.3582    & 0.4715    & 0.4523    & 0.4733    & 0.4842    & 0.4859    & \textbf{0.5115*} \\
        && N@3  & 0.2681    & 0.2764    & 0.2842    & 0.4127    & 0.3909    & 0.4126    & 0.4281    & 0.4314    & \textbf{0.4487*} \\ 
        \bottomrule
    \end{tabular}
    \caption{Comparison across datasets. * denotes our results are significantly better than the best baselines (p $<$ 0.05).}
    \label{tab:overall}
\end{table*}

\subsection{Experimental Setup}
\subsubsection{Dataset.}

For our evaluation, we use two industrial datasets of user-smart vehicle interactions and two public datasets of user-smart home interactions. The vehicle datasets contain logs from 100 randomly sampled private smart vehicles of the same model in two local cities. Each log entry includes the vehicle's ID, signal name and value, and timestamp. These datasets comprise 9.5 million events over 90 days, covering 357 sensors with 487 types of events. The sensors are categorized into comfort (129), infotainment (76), monitoring (61), safety (48), and control (43). They track everyday activities like gear changes and air conditioning adjustments. To protect privacy, vehicle IDs are anonymized using random strings, and sensitive information such as locations is omitted. An ethics board from a local institution has approved the dataset. More details can be found in~\cite{data}. The two public smart home datasets Aruba and Milan come from the WSU CASAS project, which are commonly used in the smart home activity prediction~\cite{park2022enhanced}. Similar to the vehicle data, these logs capture user-triggered events with information on sensor location, name, readings, and timestamp. The sensors monitor various household items including refrigerators, containers, lights, motion detectors, toilets, ovens, curtains, and doors. 

\setlength{\tabcolsep}{1.5mm}
\begin{table}
    \small
    \centering
    \renewcommand{\arraystretch}{1.0}    
    \begin{tabular}{ccccccccc} 
        \toprule
        \multicolumn{3}{c}{Embedding}& \multirow{2}{*}{PT} & \multicolumn{2}{c}{H@1} & \multicolumn{2}{c}{H@3}\\

        \cmidrule(lr){1-3} \cmidrule(lr){5-6} \cmidrule(lr){7-8}
        ID & Log\_Q & Log\_E& & City A & City B & City A & City B\\
        \midrule
        \checkmark&&&&0.2774&0.3056&0.4835&0.5186\\
        &\checkmark&&&0.3064&0.3298&0.5125&0.5507\\
        &&\checkmark&&0.3097&0.3354&0.5172&0.5551\\
        \checkmark&\checkmark&&&0.3140&0.3396&0.5209&0.5584\\
        \checkmark&&\checkmark&&0.3176&0.3427&0.5247&0.5623\\
        \checkmark&\checkmark&&\checkmark&0.3439&0.3722&0.5495&0.5871\\        \checkmark&&\checkmark&\checkmark&\textbf{0.3450}&\textbf{0.3743}&\textbf{0.5529}&\textbf{0.5904}\\
        \bottomrule        
    \end{tabular}
    \caption{Ablation Study. Log\_Q denotes enhanced logs generated by Qwen, and Log\_E by Ernie Bot.}   
    \label{tab:ablation}
\end{table}



\subsubsection{Evaluation Settings.}

The log data in our datasets exhibit a distinct long-tail distribution, with high-frequency signal events occurring much more frequently than low-frequency ones. Consequently, widely-used metrics in recommendation systems, like \textit{Hit Ratio} and \textit{Normalize Discounted Cumulative Gain}, become less effective as they are heavily influenced by the prediction results of high-frequency signals. Therefore, we adopt class-wise metrics, specifically class-wise Hit Rate@K (abbreviated as \textit{H@K}) and class-wise NDCG@K (\textit{N@K}). \textit{H@K} calculates the likelihood of the target event appearing in the Top-K recommendations on average for each event class while \textit{N@K} evaluates the ranking of the target event within the Top-K list. Considering practical applications like providing real-time shortcut recommendations to users, a larger $K$ could significantly increase cognitive load and reduce usability. Hence, we focus on smaller values of $K$, specifically $K\in\{1,3\}$. This choice reflects a realistic scenario where a user selects a needed function from a set of three suggested options. Ernie bot is used to enhance the logs in this experiment.

To realistically simulate the scenario of actual deployment and prevent data leakage, we divided our dataset into training, validation, and testing sets based on time. Each dataset is divided into the training set, validation set, and test set in a ratio of 2:1:7 in chronological order. For the validation and testing sets, each user's interaction logs, often comprising tens of thousands of records, were segmented into non-overlapping sequences using a sliding window of length 6. Each sequence was constructed using the first 5 events to predict the 6$^{th}$ event.

\subsubsection{Implementation Details.}

We trained our networks using the Adam optimizer~\cite{kingma2014adam}, with hyper-parameter optimization conducted on the validation set. The mini-batch size was fixed at 256, and we set the embedding size at 256 for tokens in prompts and behavior sequences. The length of interaction sequences ranged from 1 to 10. During the training phase, the learning rate was adjusted within a range of \([5e^{-3}, 5e^{-4}, 5^{e-5}, 5^{e-6}]\), and weight decay parameters were explored within \([1e^{-3}, 1e^{-4}, 1e^{-5}]\). Additionally, we experimented with different prompt token numbers \(K\), specifically tuning them within the range of \([2, 4, 6, 8, 10]\). The network training extended to a maximum of 100 epochs, a duration typically sufficient to ensure convergence. The model selection was based on the best performance observed on the validation set. All networks were developed using PyTorch and Huggingface frameworks and were run on an RTX 4090 GPU. Each experiment is repeated 20 times with different seeds.

\subsection{Overall Performance (RQ1)}
\subsubsection{Baselines.}
We compare our method against eight related baseline methods using the real-world smart vehicle dataset. These methods are divided into two categories: 1) algorithms for human activity prediction in smart space: CasLSTM~\cite{krishna2018lstm}, DnnLSTM~\cite{park2022enhanced}, ConvLMLC~\cite{wang2022predicting}, and ASGen~\cite{song2024learning}; 2) sequential recommendation algorithms applicable to scenarios such as movie watching and online shopping: BERT4Rec~\cite{sun2019bert4rec}, DuoRec~\cite{qiu2022contrastive}, CL4Rec~\cite{xie2022contrastive}, and RecFormer~\cite{li2023text}. We utilized official implementations for BERT4Rec, DuoRec, CL4Rec, and RecFormer, while for the other models, we utilized PyTorch to replicate these methods as per the specifications provided in their original paper. Our code is included in the supplementary material.

\subsubsection{Results.}
Table~\ref{tab:overall} shows that our method demonstrated superior performance on all metrics across four datasets compared to existing approaches. We attribute these improvements to the fact that our enhanced representation effectively incorporates semantic information, relationships, and timing of events for prediction, allowing for a more nuanced understanding of user behavior patterns. Furthermore, while both RecFormer and our method utilize LLM-enhanced textual information, our approach also integrates this information with personalized prompts to guide the model and make predictions tailored for the target user. This integration results in a more effective model than RecFormer. 

\begin{figure}
    \centering
    \includegraphics[width=1\linewidth]{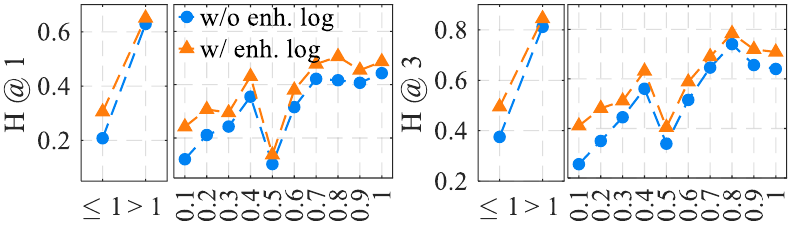}
    \caption{Impact of event frequency on prediction performance. The first and third panels categorize events into two groups based on whether their frequency exceeds 1\%. The second and fourth panels further divide events with frequencies below 1\% into ten groups using 0.1\% intervals.}
    \label{fig:long_tail}
\end{figure}

\subsection{Ablation Study (RQ2)}
\label{sec:ablation_study}
Our approach utilizes prompt tuning with two types of embeddings: ID-based and enhanced-log. The enhanced logs are generated using either the Qwen or Ernie bot. We conducted an analysis across seven experimental setups, each featuring different combinations of modules, as shown in Table~\ref{tab:ablation}. Our findings reveal that using only event IDs to understand the sequential relationships in log sequences yields a hit rate (H@1) of 0.2774 and 0.3056. Utilizing LLM-enhanced semantics textual description to create text embeddings enhances the accuracy to 0.3176 and 0.3427. Notably, the enhanced logs from both Qwen and Ernie bot exhibit high similarity, leading to comparable performance across both models. Incorporating personalized prompts further boosts performance, reaching a hit rate of 0.3450 and 0.3743 H@1. The ablation study clearly demonstrates the significant effectiveness of our LLM-enhanced descriptions and personalized prompts in enhancing predictive accuracy.

\subsection{Performance of Events over Frequencies (RQ3)}
Generally, for datasets with a significant long-tail distribution in event frequencies. Common events, due to their sufficient samples, tend to be learned more effectively compared to rare events that suffer from insufficient training data. To explore the impact of event frequency on model performance, we divided the events of the vehicle datasets into two main groups based on their occurrence in the log records: events with frequencies above 1\% and those below 1\%. For the latter, we further segmented the data into ten subgroups at 0.1\% intervals, as depicted in Fig.~\ref{fig:long_tail}. Our analysis focused on comparing the model's predictive accuracy with and without the use of LLM-enhanced descriptions (i.e., text embedding). The results (the 1$^{st}$ and 3$^{rd}$ panels in Fig.~\ref{fig:long_tail}) clearly show that for events occurring less than 1\% of the time, LLM-enhanced descriptions significantly improve the model’s predictions, with increases of over 40\% in H@1 and 30\% in H@3. Conversely, for more frequent events, the improvements, although present, were modest. Further detailed observation of the subdivided frequencies below 1\% reveals that events with frequencies under 0.2\% saw increases in \textit{H@1} and \textit{H@3} by about 0.1 or more. These findings demonstrate the effectiveness of our LLM-enhanced description for log data, especially for low-frequency events.

\begin{figure}
    \centering
    \includegraphics[width=1\linewidth]{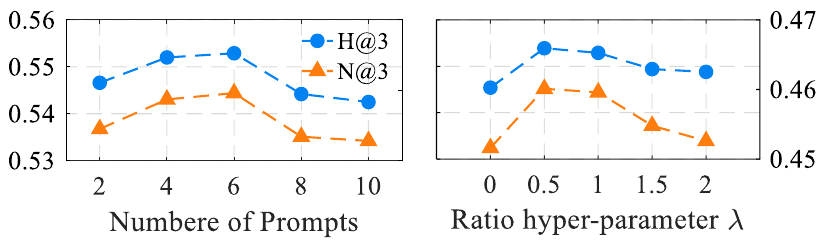}
    \caption{Impact of different hyper-parameters.}
    \label{fig:para}
\end{figure}

\subsection{Hyper-parameter Analysis (RQ4)}

With the smart vehicle datasets, we analyzed the effects of different parameters: the number of prompts and the $\lambda$ parameter used during graph construction. Fig.~\ref{fig:para} illustrates a mild increase in both H@3 and N@3 metrics as the number of prompts rises, achieving peak performance with six prompts. This improvement is likely due to the ability of additional prompts to more accurately capture user preferences in detail. In contrast, the impact of the $\lambda$ parameter was subtler. Optimal results were attained with $\lambda$ set to 0.5, as higher $\lambda$ values tend to overlook important sequential relationships, whereas lower values struggle to effectively differentiate between sequential and co-occurrence relationships. These results indicate that the model is quite robust, showing limited sensitivity to variations in these parameters.

\subsection{Discussion}

Our approach does not impose computational costs associated with LLMs during both training and inference. LLMs are only used in the preprocessing stage to enhance incomplete signals, and the enhanced descriptions are stored. During training and inference, the model simply retrieves these representations, avoiding the need for further LLM calls. This method minimizes LLM usage, limiting it to the number of unique signals, making it well-suited for large-scale log data. Furthermore, in our experiments, the inference time for a single sequence of length 10 is less than 0.1s on an NVIDIA 4090 GPU.

\section{Conclusion}

In this paper, we propose a novel approach that incorporates LLM-enhanced logs and personalized prompts. Our approach constructs a graph that captures individual behavior preferences derived from their interaction histories, which effectively transforms into a soft continuous prompt to precede the sequence of user behaviors. our approach then leverages the vast general knowledge and robust reasoning capabilities of a pretrained LLM to enrich the oversimplified and incomplete log records. Our methodology's effectiveness has been validated through rigorous evaluation on large-scale datasets involving user interactions with smart vehicles and smart homes, demonstrating the applicability of our framework in smart environment settings.

\section{Acknowledgments}
We are grateful to Professor Xiaohong Guan for his kind support of this work and anonymous reviewers for their insightful comments. This work is supported by the Science and Technology Foundation of the State Grid Corporation of China: Research on Fundamental Technologies of Power Science Computing through the Integration of Mechanism, Data, and Knowledge (No. 5700-202440332A-2-1-ZX).


\bibliography{aaai25}

\newpage

\appendix
\section{Description of the Industrial Dataset}
In this paper, we present two large-scale, real-world datasets from the industrial sector. These datasets comprise logs of human-vehicle interactions collected from 100 private smart automobiles in two different cities (50 vehicles per city), over a 90-day period through January to March. Each log entry captures key details such as the vehicle ID, signal name, signal value, and timestamp. In total, the datasets include over 9.55 million events generated by 357 types of sensors. These sensors monitor everyday interactive functions like gear-shift positions, with a full list provided in Table~\ref{tab:signals}. To ensure privacy, all vehicle IDs have been anonymized using randomly generated strings, and the datasets do not contain any sensitive information, such as location data or navigation destinations. The collection and use of these datasets have been reviewed and approved by the local institution's ethics review board. 

\begin{table}
    \centering
    \footnotesize
    \renewcommand{\arraystretch}{1.1}
    \caption{The event category in our dataset and the sensor number in each category.}    
    \label{tab:signals}
    \begin{tabular}{lll}
        \hline
        \textbf{Category} & \textbf{\#} & \textbf{Example} \\ 
        \hline
        Air Conditioner     & 51    & Heat Request              \\	
        Entertainment       & 47    & Media Volume Configuration\\	
        Energy Management   & 43    & Fuel Level Position       \\	
        Door \& Window      & 37    & Front Left Door Status    \\	
        ADAS                & 30    & APA Func Status           \\	
        Remote Request      & 29    & Remote Close Trunk Request\\	
        Seat Adjustment     & 19    & Driver Buckle Status      \\	
        Locking System      & 18    & Tank Lock Request         \\	
        Steering Wheel      & 16    & Steer Wheel Heat Request  \\  
        Signal Lighting     & 15    & Reverse Light Status      \\
        Rear Mirror         & 13    & Rear Mirror Fold Status   \\
        Gear \& Velocity    & 11    & Gear Shift Position       \\
        Tire Monitoring     & 9     & Front Left Tire Pressure  \\
        In-car Lighting     & 7     & Ambient Light Status      \\
        Window Wiper        & 3     & Window Wiper Status       \\
        Misc.               & 9     & Limp Home Status          \\        
        \hline
    \end{tabular}
\end{table}


The sensors capture both continuous and discrete events. Our goal is to convert these events into dense embeddings using one-hot vectors. However, creating an embedding for each value of a continuous event isn't practical. That approach would lead to an overwhelming number of parameters, complicating and slowing down the learning process. To avoid this issue, we group similar continuous event values into groups, or "bins". We choose the number of bins in a way that keeps values within the same bin as close as possible to each other, while making sure that values from different bins are as distinct as possible. This discretization process transforms continuous events into a manageable set of discrete states. As a result, we can effectively learn embeddings for each event-state pair.

We also observed that there exist events that almost always co-occur. An instance of this is the co-occurrence of the reverse gear status event and the reverse light status event. Due to the function logic, these two types of events should be occur at the same time. However, since the combination of such events does not offer any additional information over a single event, we replace the combination with a single event (e.g., removing all the reverse light status events) to eliminate redundancy. 

After discretizing the continuous signals, each record in our dataset includes an event's name along with its discrete state. The state of an event can significantly influence its meaning; the same event can imply different actions or intentions based on its status. For instance, ``AntiTheft\_1'' might suggest that the driver has exited the vehicle, whereas ``AntiTheft\_0" typically indicates the driver is entering the vehicle. Given this variability, it's crucial to represent events differently depending on their status.

To achieve this, we create a unique pairing of each event with its corresponding status, resulting in 487 distinct event-status combinations. Each of these pairs is then converted into a one-hot vector. These vectors serve as input to a standard PyTorch Embedding layer, which transforms them into dense, continuous embeddings. Through training, our approach refines these embeddings so that those representing similar intentions are closer together in the embedding space. For example, ``AntiTheft\_0'' would be nearer to ``DoorLock\_0'' but further from ``AntiTheft\_1''.

\begin{figure}
    \centering
    \includegraphics[width=1\linewidth]{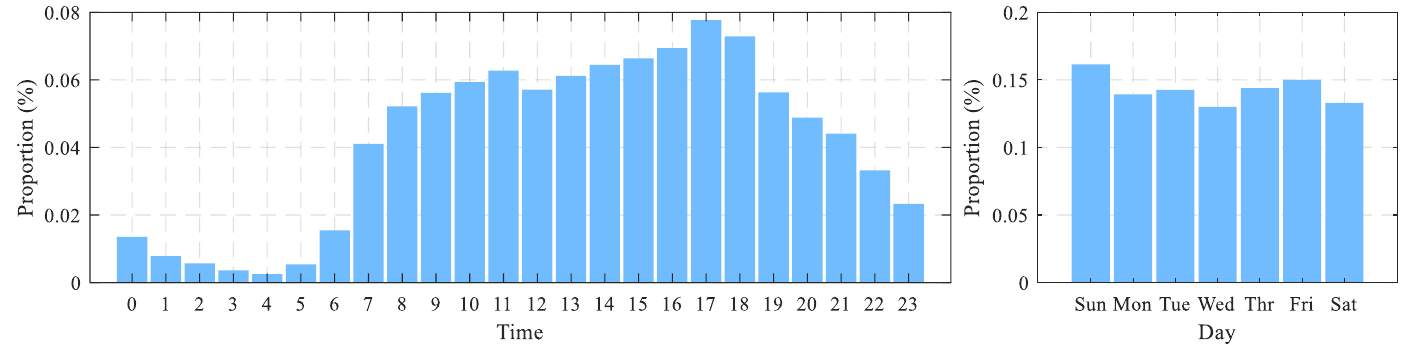}
    \caption{Event Distribution Across Time.}
    \label{fig:data_sta}
\end{figure}

In our dataset, the most frequently occurring events align closely with typical daily vehicle usage. These include events such as starting and stopping the car, shifting gears, opening and closing doors, activating the heated steering wheel, applying the brakes, adjusting windows, and toggling the seatbelt. Also, the event distribution across time of day and days of the week is displayed in Fig.~\ref{fig:data_sta}. \textbf{However, we are \textbf{not} permitted to make these datasets publicly available due to restrictions imposed by the data provider.}

\end{document}